\documentclass{elsart3}

\usepackage{graphicx}
\begin{document}
\begin{frontmatter}
\title{Peak effect at the weak- to strong pinning crossover}
\author{J.A.G.~Koopmann},
\ead{koopmann@phys.ethz.ch}
\author{V.B. Geshkenbein}, and
\author{G. Blatter}
\address{Theoretische Physik, ETH-H\"onggerberg, CH-8093 Z\"urich, Switzerland}
\begin{abstract}
In type-II superconductors, the magnetic field enters in the form of vortices;
their flow under application of a current introduces dissipation and thus
destroys the defining property of a superconductor. Vortices get immobilized by
pinning through material defects, thus resurrecting the supercurrent. In weak
collective pinning, defects compete and only fluctuations in the defect density
produce pinning. On the contrary, strong pins deform the lattice and induce
metastabilities. Here, we focus on the crossover from weak- to strong bulk
pinning, which is triggered either by increasing the strength $f_\mathrm{p}$ of
the defect potential or by decreasing the effective elasticity of the lattice
(which is parametrized by the Labusch force $f_\mathrm{Lab}$). With an
appropriate Landau expansion of the free energy we obtain a peak effect with a
sharp rise in the critical current density $j_\mathrm{c} \sim j_0 (a_0\xi^2
n_p) (\xi^2/a_0^2) (f_\mathrm{p}/f_\mathrm{Lab} -1)^2$.
\end{abstract}
\end{frontmatter}

\begin{figure}
\begin{center}
\includegraphics[scale=0.4]{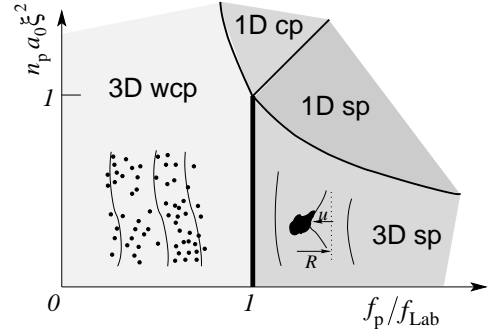}
\caption[]{Pinning diagram delineating the crossovers between various pinning
regimes involving collective versus individual pinning and 1D-line versus
3D-bulk pinning: 3D wcp -- bulk weak collective pinning, 1D cp -- collective
line pinning, 1D sp -- strong line pinning, 3D sp -- bulk strong pinning. The
crossover from weak- to strong bulk pinning occurs at the Labusch force
$f_\mathrm{Lab}$.} \label{fig:sp_wp_dia}
\end{center}
\end{figure}

Pinning of vortices by material defects is crucial in establishing the
dissipation-free flow of a supercurrent. In recent years, the focus has been on
weak collective pinning theory \cite{LO_79} describing the action of many
competing defects; pinning then is due to fluctuations. The critical current is
determined by the statistical summation of the competing pins \cite{review} and
is characterized by a quadratic dependence on the defect density. On the
contrary, first attempts to describe vortex pinning go back to Labusch
\cite{Lab_69}, who studied the action of strong individual pins. Strong pins
deform the lattice \cite{LO_NSC_86,Brandt_86gen,OvIv_91,Schoenenberger_96} and
induce metastabilities; the pinning energy landscape becomes multi-valued,
producing a non-zero average of the pinning force and a critical current which
is linear in the defect density. These two approaches have been related in a
recent work, where we have studied the pinning diagram exhibiting crossovers
between various regimes when varying the defect density $n_\mathrm{p}$ measured
with respect to the vortex density $1/a_0^2$ and the pinning force
$f_\mathrm{p}$ measured with respect to the elasticity, cf.\
Fig.~\ref{fig:sp_wp_dia}. At low defect density, vortex lattice (bulk) pinning
is relevant; for weak defect strength pinning is collective, whereas defects
with a force larger than the Labusch force $f_\mathrm{Lab}$ (see below) pin the
lattice individually. At higher defect densities we have established two
regimes where pinning acts on individual vortex lines.

In this work, we focus on the crossover from weak- to strong bulk pinning,
which is triggered either by increasing the defect strength $\kappa\sim
f_\mathrm{p}/\xi$ ($\xi$ is the coherence length of the superconductor) or by
decreasing the effective elasticity $\bar{C}$ of the lattice; the critical
defect force at the crossover is the Labusch force $f_\mathrm{Lab} =
\bar{C}\xi$. The crossover is structurally related to the Landau theory of
first order phase transitions and can be analyzed with an appropriate Landau
expansion of the free energy. As a result, we obtain a strong increase in the
critical current density at the crossover,
\begin{equation}
    j_\mathrm{c} \sim j_0 (a_0\xi^2 n_p) \frac{\xi^2}{a_0^2}
\left(\frac{f_\mathrm{p}}{f_\mathrm{Lab}} -1\right)^2, \label{sp-jc}
\end{equation}
where $j_0$ is the depairing current density and $a_0=(\Phi_0/B)^{1/2}$ is the
mean vortex spacing ($\Phi_0$ is the superconducting flux unit); see below for
the precise definition of $f_\mathrm{p}$ and $f_\mathrm{Lab}$. The crossover
from weak- to strong pinning can be triggered by reducing the effective
elasticity $\bar{C}\propto f_\mathrm{Lab}$, producing a peak in the current
density \cite{LO_79}.

Below, we will describe the strong pinning situation in more detail and use a
Landau type of expansion to find the jumps in the free energy landscape. These
jumps are then used to derive the result (\ref{sp-jc}) for the critical current
density.

In order to derive a quantitative criterion for strong pinning, we consider the
effect of a single (strong) defect on the vortex lattice. We assume a defect
located at the origin with pinning potential $e_\mathrm{p}({\bf r})$ producing
the pinning contribution
\[
    E_\mathrm{p} ({\bf r,u}) = \sum_\nu e_\mathrm{p}({\bf r})
    \delta^2({\bf R-R}_\nu -{\bf u(R}_\nu,z))
\]
to the free energy density of the vortex system; ${\bf r=(R},z)$ and the
vortices are positioned at ${\bf R}_\nu -{\bf u(R}_\nu,z)$, with ${\bf R}_\nu$
the equilibrium positions and ${\bf u}$ the displacement field. The elastic
part of the vortex lattice free energy reads \cite{review}
\begin{equation}
    \mathcal{F}_\mathrm{el} = \int \frac{d^3k}{(2\pi)^3} \, u^\alpha(-{\bf k})
    \left[ G^{\alpha\beta}({\bf k})\right]^{-1} u^\beta ({\bf k}),
\end{equation}
with $G^{\alpha\beta}({\bf k})$ the Fourier transform of the elastic Green
function $G^{\alpha\beta}({\bf r})$; Greek indices denote the in-plane
components $(x,y)$ and we sum over double indices.

The displacement field ${\bf u}$ is obtained from variation of the total energy
(elastic and pinning),
\begin{eqnarray}
    u^\alpha ( {\bf r}_\nu) &=& - \!\! \int \! d^3r' \; G^{\alpha\beta}({\bf r}_\nu-{\bf
    r}')[\partial_{u^\beta} E_\mathrm{p}]({\bf r}',{\bf u}') \label{general-u} \\
    &=& G^{\alpha\beta}({\bf R}_\nu-{\bf R}_\mathrm{d},z)
    f_{\mathrm{p}}^{\beta} [{\bf R}_\mathrm{d} + {\bf
      u(R}_\mathrm{d},0),0] , \nonumber
\end{eqnarray}
with ${\bf r}_\nu=({\bf R}_\nu,z)$ and ${\bf f}_\mathrm{p}= - \nabla_u
e_\mathrm{p}({\bf u})$ the pinning force of the defect. In the last equation,
we have assumed a defect of range much smaller than $a_0$ pinning at most one
vortex; we further have chosen ${\bf R}_\mathrm{d}$ as the distance to the
vortex closest to the defect. Evaluating (\ref{general-u}) for ${\bf
r}_\nu=({\bf R}_\mathrm{d},0)$, we arrive at the self-consistency equation
\begin{eqnarray}
     u^{\alpha} ({\bf R},0) &=& G^{\alpha\beta} ({\bf r}=0) f_\mathrm{p}^\beta
     [ {\bf R+u( R } , 0 ),0] \nonumber \\
     &= & \bar{C}^{-1} f_\mathrm{p}^\alpha[{\bf R+u(R},0),0]
    \label{self_consist_u}
\end{eqnarray}
with the effective elastic constant
\begin{equation}
\int \frac{d^3 k}{(2\pi)^3} \, G^{\alpha\beta}({\bf k}) = \bar{C}^{-1}
\delta^{\alpha\beta} .
\end{equation}
For $ a_0< \lambda$ the effective elasticity is $\bar{C}\sim
\varepsilon_0/a_0$ with $\varepsilon_0=(\Phi_0/4\pi\lambda)^2$ the
vortex line energy. The solution at (${\bf R},0$) allows for the
calculation of the whole displacement field ${\bf u(R}_\nu,z)$,
\begin{equation}
    u( {\bf R}_\nu,z) = G^{\alpha\beta}({\bf R}_\nu-{\bf R},z) \bar{C} {\bf
    u(R},0),
\end{equation}
and we can calculate the total free energy $e_\mathrm{pin}({\bf u,R})$
(containing both, the contribution from elasticity and disorder) of the vortex
system as a function of ${\bf u(R},0)$,
\begin{eqnarray}
    e_\mathrm{pin} &=&\frac{1}{2} \int \!\!\! \frac{d^3k}{(2\pi)^3}
    \big[ G^{\alpha\gamma}(-{\bf k}) \left[ G^{\alpha\beta}({\bf k})\right]^{-1}
    G^{\beta\delta}({\bf k})\nonumber \\ && \times u^\gamma({\bf R},0)
    u^\delta({\bf R},0)\big] + e_\mathrm{p}({\bf R+u({\bf R},0)}) \nonumber \\
    &=& \frac{1}{2} \bar{C} u^2  + e_\mathrm{p}({\bf R+u}) \label{eq_for_u} .
\end{eqnarray}
\begin{figure}
\begin{center}
\includegraphics[scale=0.35]{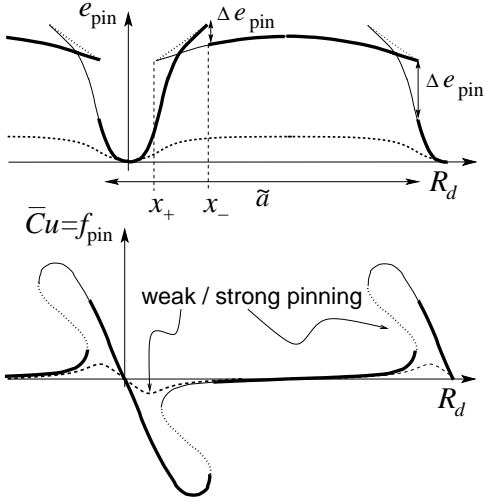}
\caption[]{The pinning energy $e_\mathrm{pin}$ and the force $f_\mathrm{pin}$
are sketched as a function of the drag $R_\mathrm{d}$ between the vortex system
and the defect. In the case of weak pinning (dashed line), the two quantities
are single-valued (hence the average over the pinning force is zero), whereas
in the strong pinning case (solid line, the dotted parts denote unstable
branches) the average over the force is determined by the jump $\Delta
e_\mathrm{pin}$ in the pinning energy. The thick lines show the choice of the
two possible branches that occurs when dragging the lattice through the defect
along the positive $x$ direction.} \label{sketch-epin-of-x}
\end{center}
\end{figure}

For weak pinning, the displacement $u$ is small and the solution
${\bf u } \approx {\bf f}_\mathrm{p}({\bf R})/\bar{C}$ of
(\ref{self_consist_u}) is unique; the pinning energy
$e_\mathrm{pin}(R)=e_\mathrm{pin}({\bf u(R),R})$ then is a
single-valued function. Strong pinning, however, produces
multi-valued solutions ${\bf u}({\bf R},0)$ and
$e_\mathrm{pin}(R)$, as can be seen in Fig.\
\ref{sketch-epin-of-x}. The displacement turns multi-valued when
$\partial_R u \to \infty$. Assuming a defect symmetric in the
plane ($e_\mathrm{p}({\bf R})=e_\mathrm{p}(R)$) and dragging the
lattice through the defect center along the x-axis (${\bf
R}_\mathrm{d}=(x,0)$), we find
\[
    \partial_x u(x) = \frac{ f'_\mathrm{p}(x+u) }{\bar{C}-f'_\mathrm{p}
    (x+u)}
\]
(note that $x>0$ implies $u<0$). The displacement collapses when the (negative)
curvature $f_\mathrm{p}'$ of the defect potential overcompensates the
elasticity of the lattice,
\begin{equation}
    \partial_x f_\mathrm{p} (x+u) = - \partial_x^2 e_\mathrm{p} (x+u) = \bar{C}
    .\label{labusch-cond}
\end{equation}
This (Labusch) criterion \cite{Lab_69} for strong pinning involves the maximal
negative curvature above the inflection point; it tests an individual isolated
pin and classifies it as weak or strong.

In a next step, we calculate the average pinning force in the strong pinning
case. Using the self-consistency equation (\ref{self_consist_u}), we calculate
the derivative of the pinning energy $e_\mathrm{pin}(x,y) =e_\mathrm{pin}({\bf
u(R),R})$ with respect to the drag parameter $x$,
\begin{eqnarray}
    \partial_x e_\mathrm{pin}(x,y) &=& \bar{C} {\bf u}
    \partial_x {\bf u} -
    {\bf f}_\mathrm{p}({\bf R+u}) [\hat{\bf x}+\partial_x {\bf u}] \nonumber \\
    &=& -f_\mathrm{p}^x({\bf R+u}) ;
\end{eqnarray}
this equation relates the force along $x$ exerted by the pin at
the vortex position ${\bf R + u(R},0)$ to the free energy
landscape containing both, the contribution from pinning and
elasticity. The pinning force has to be averaged over defect
locations and, due to the bistabilities, depends on the
preparation of the system; here, we focus on the critical current
density and therefore search for the maximal force against drag.
Averaging the accumulated drag force over the `impact parameter'
$y$, we obtain
\begin{eqnarray}
    \langle f_\mathrm{pin} \rangle &=& \int_0^{L_x} d x \int_0^{L_y} d y\;
    \frac{f_\mathrm{p}^x ({\bf R + u})}{L_x L_y} \\
    &=& \int_0^{L_x} d x \int_0^{L_y} d y \; \frac{-\partial_x
e_\mathrm{pin}(x,y)}{L_x L_y} ;
    \nonumber
\end{eqnarray}
the integral over $x$ can equally well be interpreted as an integrated drag
force or as an average over pin locations (maximized with respect to the
bistable branches). We express the average along the $x$-axis through the jump
$\Delta e_\mathrm{pin}(y)$ in the pinning energy,
\begin{equation}
    \langle f_\mathrm{pin} \rangle =  -\int_{-a_0/2}^{a_0/2} \frac{d y}{a_0}
    \frac{\Delta e_\mathrm{pin}(y)}{\tilde a(y)} = - \frac{t_\perp}{a_0^2} \Delta
    e_\mathrm{pin} , \label{fpin}
\end{equation}
where we have assumed a maximal trapping distance $t_\perp$ along the $y$ axis
and require the pin not to overdrag the vortex; then the periodicity of
$e_\mathrm{pin}$ is the same as the lattice periodicity and $\tilde a=a_0$. The
remaining task is the determination of the jump $\Delta e_\mathrm{pin}$; as we
will see below, it does not depend on the impact parameter $y$.

\begin{figure}
\begin{center}
\includegraphics[scale=0.2]{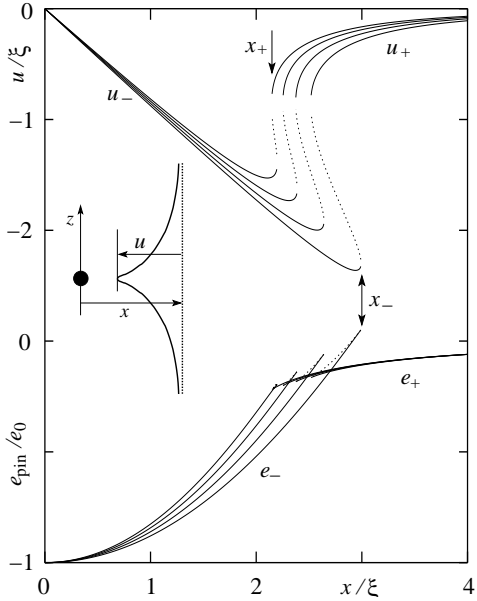}
\caption[]{Displacement $u$ and energy $e_\mathrm{pin}$ versus drag $x$ of the
vortex lattice relative to a defect producing a Lorentzian pinning potential
$e_\mathrm{p,L} = -e_0/(1+u^2/\xi^2)$ (we measure units of length in $\xi$ and
units of energy in $e_0$). The evolution of the bistability appearing above the
Labusch criterium is shown for pinning centers of increasing strength with
$\bar{C}/\kappa = 0.85, 0.75, 0.65, 0.55$; dashed lines denote unstable
branches. The inset shows the geometry defining the drag parameter $x$ and the
displacement $u$.} \label{fig:strong_pin}
\end{center}
\end{figure}

In a first step, we discuss the solutions of the self-consistency equation
(\ref{self_consist_u}) when dragging the lattice through the defect center
along the positive $x$ direction and calculate the displacement $u$ numerically
for a Lorentzian defect potential
\[
    e_\mathrm{p,L}(u)= -\frac{e_0}{1+u^2/\xi^2} .
\]
The result is shown in Fig.~\ref{fig:strong_pin}, where we have plotted the
displacement $u$ and the pinning energy $e_\mathrm{pin}$ for different values
of $\bar{C}/\kappa \sim f_\mathrm{Lab}/f_\mathrm{p}$, where
$-\kappa=-e_0/2\xi^2$ is the maximal negative curvature of the defect potential
(realized at $u=u_\kappa=\xi$), which is the relevant curvature in the Labusch
criterion (\ref{labusch-cond}). The branch $u_-$ ($u_+$) corresponds to the
pinned (unpinned) vortex and becomes unstable at $x_-$ ($x_+$); the dashed
lines denote unstable branches.

\begin{figure}[t]
\begin{center}
\includegraphics[scale=0.33]{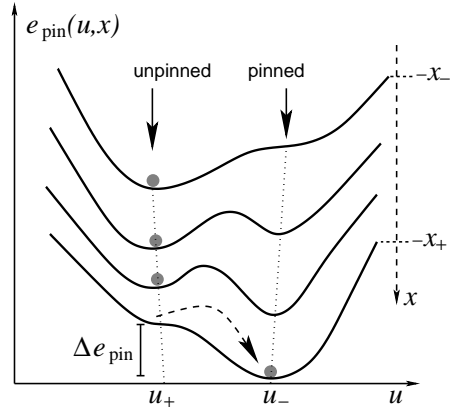}
\caption[]{Schematic plot of the pinning energy as a function of the
displacement $u$ for different (negative) values of the drag $x$. The sketch
shows the trapping process of an unpinned vortex at the point of instability
$-x_+$ and the appearance of the jump $\Delta e_\mathrm{pin}$.}
\label{sketch-epin-of-u}
\end{center}
\end{figure}

The solutions $u(x)$ of the self-consistency equation
(\ref{self_consist_u}) correspond to the minima of the functional
$e_\mathrm{pin}[u,x]$ for fixed $x$, cf.\
Fig.~\ref{sketch-epin-of-u}. Starting far from the defect with a
negative drag parameter $x$, there is only one solution $u_+$
corresponding to the unpinned vortex system, cf.\ Figs.\
\ref{sketch-epin-of-x} and \ref{sketch-epin-of-u}. With the vortex
approaching the defect a second minimum appears, but is separated
from the first one by a barrier. The relevant solution therefore
remains the unpinned one, although at some point the pinned
solution becomes lower in energy. For $x=-x_+$ the unpinned
solution becomes unstable and the vortex gets trapped, as shown in
Fig.~\ref{sketch-epin-of-u}. It remains so until at $x=x_-$ the
pinned solution $u_-$ turns unstable and the defect releases the
vortex. The corresponding jumps in the free energy are shown in
Fig.~\ref{sketch-epin-of-x}.

In order to attain a more quantitative description of the different branches
and their corresponding jumps, we perform a Landau type expansion of the free
energy close to the Labusch condition $\kappa = \bar{C}$ and use the analogy to
the Landau theory of first order phase transitions. We denote the location of
the maximal negative curvature $-\kappa$ by $u_\kappa$ and expand the pinning
potential $e_\mathrm{p}(x+u)$ around this point,
\begin{eqnarray}
    && e_\mathrm{p}(x+u)  \approx  -\epsilon + \nu (x+u-u_\kappa)\\
    &&\qquad - \frac{\kappa}{2} (x+u-u_\kappa)^2
    + \frac{\alpha}{24} (x+u-u_\kappa)^4.
    \nonumber
\end{eqnarray}
For the Lorentzian potential the expansion parameters read $u_\kappa=\xi$,
$\epsilon = e_0/2$, $\nu = e_0/2\xi$, $\kappa = e_0/2\xi^2$, and
$\alpha=3e_0/\xi^4$. We insert this expansion into the pinning energy
$e_\mathrm{pin}=Cu^2/2 + e_\mathrm{p}(R+u)$; the latter maps to the free energy
of a one-component (Ising) magnet in a magnetic field \cite{LL} if we define
the order parameter $\phi=x+u-u_\kappa$, the reduced `temperature' $\tau =
\bar{C}-\kappa$, and the `magnetic field' $h=\bar{C} (x-u_\kappa-\nu/\bar{C})$,
\begin{equation}
    e_\mathrm{pin}[u,x]= e_\mathrm{mag}[\phi,h] = \frac{\tau}{2} \phi^2 +
    \frac{\alpha}{24}\phi^4 - h \phi , \label{epin-expanded}
\end{equation}
where we have dropped the constant term.
%
%\begin{eqnarray}
%    e_\mathrm{pin}[u,x] &\approx &\frac{\bar{C}}{2} u^2 - \epsilon + \nu
%    (u+x-u_\kappa)\label{epin-expanded} \\
%    && -\frac{\kappa}{2} (u+x-u_\kappa)^2 + \frac{\alpha}{24}(u+x-u_\kappa)^4 .
%    \nonumber
%\end{eqnarray}
%%
Consider the zero field case first: In the paramagnetic phase at
high-temperatures $\tau>0$ the order parameter $\phi$ vanishes,
while $\phi \propto \pm |\tau|^{1/2}$ in the low temperature
($\tau<0$) ferromagnetic phase. The two ferromagnetic states
($\phi>0$ and $\phi<0$) are separated by an energy barrier
increasing as $\Delta e_\mathrm{mag}\propto \tau^2$. In the low
temperature phase $\tau<0$, a field $h$ induces a first order
transition across $h=0$ between the two ferromagnetic states.
Metastable/hysteretic behavior appears within the field regime
$|h|<h^\star$, with $h^\star \propto |\tau|^{3/2}$ following from
comparing the barrier $\Delta e_\mathrm{mag}$ and the field energy
$h\phi$.

The above considerations translate to the pinning problem in the
following way: the high `temperature' phase $\tau>0$ describes
weak pinning. The two low temperature states stand for the pinned
($\phi<0$) and unpinned ($\phi>0$) configurations, which transform
into one another via the first order transition. The hysteretic
regime translates to the bistable pinning domain bounded by
\[
    x_\pm = u_\kappa +(\nu \mp h^\star)/\bar{C},
\]
with $h^\star = (2/3\bar{C})\sqrt{2/\alpha} |\tau|^{3/2}$. At the instability
$x=x_-$ ($h=h^\star$) the order parameter jumps from the pinned solution $u_-=
\phi_- + u_\kappa -x_-$ to the unpinned state $u_+ = \phi_+ +u_\kappa -x_-$,
where $\phi_-= -\sqrt{2/\alpha}\, |\tau|^{1/2}$ and $\phi_+=2\sqrt{2/\alpha}\,
|\tau|^{1/2}$. As a result, we find the jump $\Delta e =
e_\mathrm{pin}(u_-,x_-) - e_\mathrm{pin} (u_+,x_-)=(9/2\alpha)\, \tau^2 $ in
the free energy; a jump of equal magnitude is found at $x_+$. The total jump
$\Delta e_\mathrm{pin}$ in the pinning energy then reads
\[
    \Delta e_\mathrm{pin} = 2 \Delta e = (9/\alpha)\, (\bar{C}-\kappa)^2 .
\]
Within this Landau description, the Labusch condition
$\bar{C}=\kappa$ corresponds to the critical end-point $T=T_c$
terminating a line of first order transitions.

\begin{figure}[h]
\begin{center}
\includegraphics[scale=0.37]{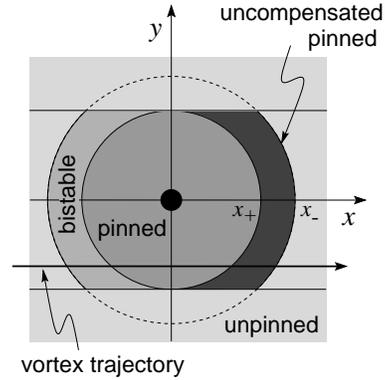}
\caption[]{Trapping area around the defect for a circular symmetric
  situation with pinned, unpinned, and bistable regions; the dark
  uncompensated area produces the net pinning force. A vortex trajectory
  for finite impact parameter $y<0$ is shown.} \label{fig:trap-geom}
\end{center}
\end{figure}

Going beyond this one-dimensional analysis, we have to account for
a finite `impact parameter' $y$ of the vortex with respect to the
defect (${\bf R}_\mathrm{d}=(x,y)$) and determine the transverse
trapping distance $t_\perp$ for the vortex to get trapped when
dragged along $x$, cf.\ Fig.\ \ref{fig:trap-geom}. Making use of
the rotational symmetry of the problem, we find circles with
radius $x_\pm$ limiting the bistable regions. The jumps then occur
when a solution is dragged through its instability line: The
unpinned vortex gets pinned when $x=-(x_+^2-y^2)^{1/2}$; this only
happens for impact parameters $|y|<x_+$, which leads to a trapping
distance $t_\perp = 2x_+$ (note, that $t_\perp \approx u_\kappa
+\nu/\bar{C}$ remains finite at the transition to weak pinning).
The pinned vortex is released when $x = (x_-^2-y^2)^{1/2}$. Note,
that the corresponding magnitudes of the jumps do not depend on
$y$ because of the planar symmetry.

We insert the value for the jump $\Delta e_\mathrm{pin}$ into (\ref{fpin}), use
$t_\perp = 2x_+$ and obtain the pinning force $\langle f_\mathrm{pin}\rangle =
-2x_+ \Delta e_\mathrm{pin}/a_0^2$. Adding the contributions of independent
strong defects with density $n_\mathrm{p}$, the total pinning force density
reads $F_\mathrm{pin}=n_\mathrm{p} \langle f_\mathrm{pin} \rangle$ and the
critical current density is obtained as $j_\mathrm{c} \sim -
cF_\mathrm{pin}/B$. Close to the Labusch condition, we finally obtain the
critical current density
\begin{equation}
    j_\mathrm{c} \sim j_0 \; a_0\xi^2 n_\mathrm{p} \frac{\xi^2}{a_0^2} \left(
    \frac{f_\mathrm{p}}{f_\mathrm{Lab}}-1 \right)^2 ,
\end{equation}
where we have used a defect size $\xi$, implying $t_\perp \sim \xi$; we define
the relevant defect strength $f_p = \xi \kappa =\xi \max_u
[f_\mathrm{p}^\prime(u)]$ in terms of the maximal curvature of the defect
potential. The Labusch force then takes the form $f_\mathrm{Lab}=\xi\bar{C}
\sim \varepsilon_0 \xi/a_0$. Interpolating with the weak collective pinning
result \cite{review}
\[
     j_\mathrm{c}^\mathrm{wcp}
     \sim j_0 (a_0\xi^2 n_\mathrm{p})^2
    \frac{\xi^2}{\lambda^2}
    \left(\frac{f_\mathrm{p}}{f_\mathrm{Lab}}\right)^4
\]
(we assume pinned bundles of size larger than $\lambda$), we observe a sharp
rise in the critical current density once the strong pinning force overcomes
the weak pinning result; this peak effect \cite{LO_79} occurs close to the
Labusch condition at $f_\mathrm{p}/f_\mathrm{Lab}=1+(a_0/\lambda) \sqrt{a_0
\xi^2 n_\mathrm{p}}$.

The crossover from weak- to strong pinning can be realized in experiments:
increasing the magnetic field and approaching the upper critical field
$H_{c_2}$ leads to a marked softening of the elastic moduli, thereby reducing
$f_\mathrm{Lab}\propto \bar{C}$; the resulting crossover into the strong
pinning regime then is observed as a peak \cite{LO_79,pippard1969} in the
current density. Whether such a crossover from weak- to strong pinning can
explain the recently observed peak effects (see, e.g.,
\cite{marchevsky2110,avraham2001} and Refs.\ therein) remains to be seen.

We acknowledge discussions with Anatoly Larkin and financial support from the
Swiss National Foundation.


\begin{thebibliography}{99}

\bibitem{LO_79} A.I.\ Larkin and Yu.N.\ Ovchinnikov,
   J.\ Low Temp.\ Phys. \textbf{34}, 409 (1979).

\bibitem{review} G.\ Blatter, M.V.\ Feigel'man, V.B.\ Geshkenbein,
   A.I.\ Larkin, and V.M.\ Vinokur,
   Rev.\ Mod.\ Phys.\ \textbf{66}, 1125 (1994).

\bibitem{Lab_69} R.\ Labusch,
   Cryst.\ Latt.\ Defects \textbf{1}, 1 (1969).

\bibitem{LO_NSC_86} A.I. Larkin and Yu.N. Ovchinnikov,
  in {\itshape Nonequilibrium Superconductivity},
  eds.\ D.N.\ Langenberg and A.I.\ Larkin
  (Elsevier Science Publishers, 1986), p. 493.

\bibitem{Brandt_86gen} E.H.\ Brandt,
   Phys.\ Rev.\ B \textbf{34}, 6514 (1986).

\bibitem{OvIv_91} Yu.N.\ Ovchinnikov and B.I.\ Ivlev,
   Phys.\ Rev.\ B \textbf{43}, 8024 (1991).

\bibitem{Schoenenberger_96} A.\ Sch\"onenberger, A.I.\ Larkin,
   E.\ Heeb, V.B.\ Geshkenbein, and G.\ Blatter,
   Phys.\ Rev.\ Lett.\ \textbf{77}, 4636 (1996).

\bibitem{LL} L.D.\ Landau and E.M.\ Lifschitz,
   \emph{Statistical Physics}, Vol.\ \textbf{5}
   (Pergamon Press, London/Paris, 1958).

\bibitem{pippard1969} A.B.\ Pippard,
    Phil.\ Mag.\ {\bf 19}, 217 (1969).

\bibitem{marchevsky2110} M.\ Marchevsky, M.J.\ Higgins, and S.\ Bhattacharya,
    Nature \textbf{409}, 591 (2001).

\bibitem{avraham2001}  M.\ Avraham, B.\ Khaykovich, Y.\ Myasoedov,
   M.\ Rappaport, H.\ Shtrikman, D.E.\ Feldman, T.\ Tamegai,
   P.H.\ Kes, M.\ Li, M.\ Konczykowski, K.\ van der Beek,
   and E.\ Zeldov,
   Nature \textbf{411}, 451 (2001).

\end{thebibliography}
\end{document}